\documentclass{article}

\usepackage{units}
\usepackage{color}
\usepackage{graphicx}
\usepackage{hyperref}
\usepackage[round]{natbib}

\def\abs#1{\left|#1\right|}
\def\rmE{\mathrm{E}}
\def\vecstyle{\mathit}
\def\matstyle{\mathbf}

\begin{document}

\title{Practical Realization of Bessel's Correction for a Bias-Free Estimation of the Auto-Covariance and the Cross-Covariance Functions}
\author{Holger Nobach}
\date{May 29, 2017}

\maketitle

\vspace{-0.5cm}
\begin{center}
(updated on January 16, 2020)\\
(updated on October 7, 2021)
\end{center}
\vspace{0.5cm}

\begin{abstract}
To derive the auto-covariance function from a sampled and time-limited signal
or the cross-covariance function from two such signals, the mean values must be 
estimated and removed from the signals. If no {\it a priori} information about the 
correct mean values is available and the mean values must be derived from the 
time series themselves, the estimates will be biased. For the estimation of the 
variance from independent data the appropriate correction is widely known as
Bessel's correction. Similar corrections for the auto-covariance and for the 
cross-covariance functions are shown here, including individual weighting
of the samples. The corrected estimates then can be used to correct also 
the variance estimate in the case of correlated data.
The programs used here are available online at \url{http://sigproc.nambis.de/programs}.
\end{abstract}


\section{Introduction}

The processing of measured data often requires mean-free data sets to emphasize the dynamic
characteristics of the observed process. Since the mean value often is unknown beforehand,
the standard procedure is to estimate the mean value from the measured data set
and then remove this estimated mean value from the measured values before further data processing.
For the following investigations a set of $N$ measured data samples $x_i, i=0\ldots N-1$, taken at 
their measurement times $t_i=i\Delta t$ with the regular sampling interval $\Delta t$ is assumed.
The samples can have individual weights $w_i$, which can be used to correct systematic errors
due to an askance distribution of the data values or to mask invalid data samples.
The estimate of the mean value from the available data samples then looks 
\begin{equation}
\bar x=\frac{\sum\limits_{i=0}^{N-1} w_i x_i}{\sum\limits_{i=0}^{N-1} w_i},
\label{eq:mean}
\end{equation}
which then is subtracted from all samples, yielding the new, mean-free samples $\tilde x_i=x_i-\bar x$
taken for the following data analysis. Higher-order trend removal, outliers or superimposed noise are not investigated here.

Let the mean estimator have the estimation variance $\sigma_{\bar x}^2$.
Since the variance of a sum of correlated variables is the sum of all pair-wise covariances, 
the variance of the mean estimator is\footnote{For all weights being constant, the expression reduces to 
\[\sigma_{\bar x}^2=\frac{1}{N^2}\sum\limits_{k=-(N-1)}^{N-1} \left(N-\abs{k}\right) C_k.\]}
\begin{equation}
\sigma_{\bar x}^2=\frac{\sum\limits_{i=0}^{N-1}\sum\limits_{j=0}^{N-1} w_i w_j C_{j-i}}{\left(\sum\limits_{i=0}^{N-1}w_i\right)^2},
\label{eq:meanvari}
\end{equation}
involving the unknown true auto-covariance function $C$.

If the variance of the data set is obtained from the mean-subtracted values $\tilde x_i$ as
\begin{equation}
s^2=\frac{\sum\limits_{i=0}^{N-1} w_i \tilde x_i^2}{\sum\limits_{i=0}^{N-1} w_i},
\label{eq:sq}
\end{equation}
then this estimate will have a systematic error due to the fact that the estimation 
of the mean value before with its estimation variance $\sigma_{\bar x}^2$
will reduce the remaining power in the investigated data sequence after removing the estimated mean.

The expectation of the variance estimation with the estimated mean subtracted from the data samples is
\begin{equation}
\rmE\{s^2\}=\sigma_x^2-\frac{\sum\limits_{i=0}^{N-1}\sum\limits_{j=0}^{N-1}w_i w_jC_{j-i}}{\left(\sum\limits_{i=0}^{N-1} w_i\right)^2}
\label{eq:esq}
\end{equation}
with the true variance $\sigma_x^2$ of the data and again with the true auto-covariance function $C$.
The deviation from the correct variance is exactly the variance of the mean estimator $\sigma_{\bar x}^2$.

If the variance of the mean estimation is known beforehand, then a bias-free estimate of the data variance is 
\begin{equation}
\hat s^2=s^2+\sigma_{\bar x}^2.
\label{eq:varicorrect}
\end{equation}

For $N$ independent data samples $x_i$ with their weights $w_i$, the variance of the mean estimation can be predicted as 
\begin{equation}
\sigma_{\bar x}^2=\frac{\sum\limits_{i=0}^{N-1} w_i^2}{\left(\sum\limits_{i=0}^{N-1}w_i\right)^2}\cdot \sigma_x^2.
\end{equation}
Requesting that the variance estimate $\hat s^2$ becomes bias free 
without knowing the true variance $\sigma_x^2$ beforehand leads to the estimate
\begin{equation}
\hat s^2=\frac{\sum\limits_{i=0}^{N-1} w_i}{\left(\sum\limits_{i=0}^{N-1} w_i\right)^2-\sum\limits_{i=0}^{N-1} w_i^2}\cdot \sum\limits_{i=0}^{N-1} w_i \tilde x_i^2.
\end{equation}
For all weights being constant (including that the samples are independent) 
this reduces to the expression
\begin{equation}
\hat s^2=\frac{1}{N-1}\sum\limits_{i=0}^{N-1} \tilde x_i^2,
\end{equation}
where the division by $N-1$ instead of $N$
is widely known as Bessel's correction for the variance estimate for independent data
samples, even if it is more likely attributed to Gauss \citep[p.\ 125]{kenney_keeping_51}.
Similar corrections can be made to estimates of the auto-covariance function or the cross-covariance function 
derived from two different data sets. Unfortunately, this requires considering that the data samples 
are correlated --- why one would otherwise calculate the covariance function?

It seems that in the past not much research has been made to investigate or solve this particular problem,
even if it seems to be a logical step. A literature research reflects the low interest by no appropriate 
articles in the past decades. 
The more surprising it was, that very recently a paper was published by \citet{vogelsang_yang_16}, 
using exactly the here proposed idea of deriving a prediction matrix, mapping the true covariance function
onto the expectation of the estimated one and using the inverse of this matrix to obtain a corrected
covariance function from the estimated one.
Considering this coincidence, the notation of the matrix has been adjusted accordingly 
and the title also takes this into account by introducing now a ``practical realization'' of the method.
Otherwise, the present article uses its own derivations. Different to \citet{vogelsang_yang_16},
here weighted averages are used in the estimation of the statistical properties. 
Furthermore, the investigations have been extended to the case of estimating 
the cross-covariance function between two data sets. 
Note, that in the present derivations, the primary covariance estimates are based on the normalization
considering the decreasing overlap of the observed signals for increasing lag time instead of a constant normalization factor.
Furthermore, the two-sided (symmetric) auto-covariance function is used instead of the one-sided,
because this better corresponds to the cross-covariance function and it may accelerate the computation 
by allowing the usage of the fast Fourier transform.
Finally, the bias-corrected estimation of the covariance function can be used to 
obtain an appropriate correction of the variance estimate under the condition of 
correlated data samples.

The following sections introduce the procedures to derive bias-free estimates of the auto- and the cross-covariance function
from equidistantly sampled, time-limited data sets, where the mean values are derived and subtracted from 
the data as described above. All required quantities are derived directly from the observed data. 
No further {\em a priori} information is needed.
The programs used here are available online at \url{http://sigproc.nambis.de/programs}.

\section{Auto-covariance case}

The auto-covariance $C_k$ of a data sequence, at the time instance $\tau_k=k\Delta t$, is defined as
\begin{equation}
C_k=\left\langle (x_i-\mu) (x_{i+k}-\mu)\right\rangle
\end{equation}
with the true mean value $\mu$ and the expectation $\langle\cdot\rangle$. 
Assuming a data set of $N$ samples $\tilde x_i, i=0\ldots N-1$ after removing the estimated mean value $\bar x$,
measured at time instances $t_i=i\Delta t$ and appropriate individual weights $w_i$, an estimator of the 
auto-covariance function of an aperiodic signal could look like
\begin{equation}
c_k=\frac{\sum\limits_{i=\max(0,-k)}^{\min(N,N-k)-1} w_i w_{i+k} \tilde x_i \tilde x_{i+k}}{\sum\limits_{i=\max(0,-k)}^{\min(N,N-k)-1} w_i w_{i+k}}=\frac{X_k}{Y_k}.
\label{eq:ck}
\end{equation}
Assuming a zero padding of $N$ concatenated zeros, the appropriate sums in the numerator (${\vecstyle{X}}$)
and in the denominator (${\vecstyle{Y}}$) can also be calculated by means of the (fast) discrete Fourier transform (FFT) and its inverse (IFFT) as
\begin{eqnarray}
{\vecstyle{X}}&=&\mathrm{IFFT}\left\{\abs{\mathrm{FFT}\left\{w'_i \tilde x'_i\right\}}^2\right\}\\
{\vecstyle{Y}}&=&\mathrm{IFFT}\left\{\abs{\mathrm{FFT}\left\{w'_i\right\}}^2\right\},
\end{eqnarray}
where $\left\{w'_i \tilde x'_i\right\}$ and $\left\{w'_i\right\}$ 
are the zero-padded sets of weighted data values (after mean removal)
and that of the weights respectively.

This estimator has a similar systematic error as the variance estimator above (see example in Fig.~\ref{fig:sim}b).
An appropriate estimation of the expectation of the covariance function is
\begin{equation}
\rmE\{c_k\}=C_k+\varepsilon_k,
\label{eq:eck}
\end{equation}
with the true auto-covariance function $C_k$ at lag time $\tau_k$ and the bias
\begin{equation}
\varepsilon_k=\frac{\sum\limits_{i=0}^{N-1}\sum\limits_{j=0}^{N-1}w_i w_jC_{j-i}}{\left(\sum\limits_{i=0}^{N-1} w_i\right)^2}
-\frac{\sum\limits_{i=\max(0,-k)}^{\min(N,N-k)-1}\sum\limits_{j=0}^{N-1}w_i w_{i+k} w_j(C_{j-i}+C_{i+k-j})}
{\left(\sum\limits_{i=\max(0,-k)}^{\min(N,N-k)-1}w_i w_{i+k}\right)\left(\sum\limits_{i=0}^{N-1} w_i\right)},
\label{eq:eek}
\end{equation}
which is constant for uncorrelated data, otherwise it varies with $k$.
The first term again is the variance $\sigma_{\bar x}^2$ of the mean estimator.
Since the true covariance function ${\vecstyle{C}}$ is unknown in real measurements, the prediction 
cannot be made directly. However, the relation between the true covariance function
and its estimate is linear. 
Therefore, one can built a matrix\footnote{The notation has been chosen with respect to \citet{vogelsang_yang_16}.}
${\matstyle{A}}$, mapping a hypothetical covariance function ${\vecstyle{C}}$ onto the estimated one ${\vecstyle{c}}$. 
\begin{equation}
\rmE\{{\vecstyle{c}}\}={\matstyle{A}} {\vecstyle{C}},
\end{equation}
If the matrix ${\matstyle{A}}$ has the elements $a_{kj}$ then the prediction of the estimated covariance at lag time $\tau_k$ is
\begin{equation}
\rmE\{c_k\}=\sum_{j=K_1}^{K_2}a_{kj} C_j.
\end{equation}
The range $K_1\ldots K_2$ of covariances considered should include the full range of occurring correlations, such that 
all true covariance outside this interval can be neglected.

The elements of this matrix are\footnote{If all $w_i$ are constant, then the elements of this matrix become
\[a_{kj}=\delta_{k-j}-2\frac{N-\max[\abs{j},\abs{k},\min(N,\abs{k-j})]}{N(N-\abs{k})}+\frac{N-\abs{j}}{N^2}\quad \abs{j},\abs{k}<N.\]}
\begin{eqnarray}
a_{kj}&=&\delta_{k-j}
-\frac{\sum\limits_{i=\max(0,-j,-k)}^{\min(N,N-j,N-k)-1} w_i w_{i+j} w_{i+k}}{\left(\sum\limits_{i=\max(0,-k)}^{\min(N,N-k)-1} w_i w_{i+k}\right) \left(\sum\limits_{i=0}^{N-1} w_i\right)}\nonumber\\
&&-\frac{\sum\limits_{i=\max(0,-j,k-j)}^{\min(N,N-j,N+k-j)-1} w_i w_{i+j} w_{i+j-k}}{\left(\sum\limits_{i=\max(0,-k)}^{\min(N,N-k)-1} w_i w_{i+k}\right) \left(\sum\limits_{i=0}^{N-1} w_i\right)}+\frac{\sum\limits_{i=\max(0,-j)}^{\min(N,N-j)-1} w_i w_{i+j}}{\left(\sum\limits_{i=0}^{N-1} w_i\right)^2}
\end{eqnarray}
with
\begin{equation}
\delta_i=\left\{\begin{array}{ll}1&\mbox{for $i=0$}\\0&\mbox{otherwise}\end{array}\right.
\end{equation}
or
\begin{equation}
a_{kj}=\delta_{k-j}+\frac{Y_j}{\left(\sum\limits_{i=0}^{N-1} w_i\right)^2}-\frac{G_{kj}+H_{kj}}{Y_k \left(\sum\limits_{i=0}^{N-1} w_i\right)}
\end{equation}
with
\begin{eqnarray}
{\vecstyle{G}}_k&=&\mathrm{IFFT}\left\{\mathrm{FFT}\left\{w'_iw'_{i+k} \right\}^\ast \mathrm{FFT}\left\{w'_i \right\}   \right\}\\
{\vecstyle{H}}_k&=&\mathrm{IFFT}\left\{\mathrm{FFT}\left\{w'_i \right\}^\ast \mathrm{FFT}\left\{w'_iw'_{i-k} \right\}    \right\},
\end{eqnarray}
with the conjugate complex $\cdot ^\ast$, involving again the (fast) discrete Fourier transform (FFT) and its inverse (IFFT).

The inverse of the matrix ${\matstyle{A}}^{-1}$ applied to the estimate ${\vecstyle{c}}$ yields an improved, 
bias-free estimate $\hat c$ of the covariance
\begin{equation}
\hat {\vecstyle{c}}={\matstyle{A}}^{-1} {\vecstyle{c}}.
\end{equation}

For given $N$ samples $x_i$, the covariance function after zero padding has $2N-1$ 
non-zero values $c_k$ in the range $-(N-1)\ldots N-1$. 
Unfortunately, the appropriate matrix ${\matstyle{A}}$ then has some linear dependent equations and a direct 
inverse cannot be calculated. The inverse can be calculated only, if the covariance function is limited to 
the range $K_1\ldots K_2$ with $-(N-1)<K_1\le K_2<N-1$. The improved covariance estimate then is bias free, as long as 
the true covariance of the original signal is zero outside the reduced interval of lag times $\tau_{K_1}\ldots \tau_{K_2}$. 
This coincides with the requirement that the interval of investigated lag times is larger than the longest correlation lasts 
and the observation interval of the signal is at least a little longer than the largest lag time investigated.

The improved estimate $\hat {\vecstyle{c}}$ of the covariance function then can be used to derive the
estimation variance of the mean estimator $\sigma_{\bar x}^2$ following Eq.~(\ref{eq:meanvari}),
where the true covariance ${\vecstyle{C}}$ is replaced by the improved estimate $\hat {\vecstyle{c}}$,
and finally to improve the variance estimation $\hat s^2$ following Eq.~(\ref{eq:varicorrect}).

\section{Cross-covariance case}

The cross-covariance $C_k$ of two data sequences $x_{1,i}$ and $x_{2,i}$, at the time instance $\tau_k=k\Delta t$,
is defined as
\begin{equation}
C_k=\left\langle (x_{1,i}-\mu_1) (x_{2,i+k}-\mu_2)\right\rangle
\end{equation}
with the true mean values $\mu_1$ and $\mu_2$ and the expectation $\langle\cdot\rangle$. 
Assuming data sets of $N_1$ samples $\tilde x_{1,i}, i=0\ldots N_1-1$ and $N_2$ samples $\tilde x_{2,i}, i=0\ldots N_2-1$
after removing the estimated mean values $\bar x_1$ and $\bar x_2$,
measured at time instances $t_i=i\Delta t$ and appropriate individual weights $w_{1,i}$ and $w_{2,i}$, an estimator of the 
cross-covariance function of an aperiodic signal could look like
\begin{equation}
c_k=\frac{\sum\limits_{i=\max(0,-k)}^{\min(N_1,N_2-k)-1} w_{1,i} w_{2,i+k} \tilde x_{1,i} \tilde x_{2,i+k}}{\sum\limits_{i=\max(0,-k)}^{\min(N_1,N_2-k)-1} w_{1,i} w_{2,i+k}}
=\frac{X_k}{Y_k}.
\label{eq:ckx}
\end{equation}
Assuming a zero padding of $N_2$ concatenated zeros to the sequence $x_{1,i}$ and $N_1$ concatenated zeros to the sequence $x_{2,i}$, 
the appropriate sums in the numerator (${\vecstyle{X}}$) and in the denominator (${\vecstyle{Y}}$) can also be calculated by means of the (fast) discrete Fourier transform as
\begin{eqnarray}
{\vecstyle{X}}&=&\mathrm{IFFT}\left\{\mathrm{FFT}\left\{w'_{1,i} \tilde x'_{1,i}\right\}^\ast\mathrm{FFT}\left\{w'_{2,i} \tilde x'_{2,i}\right\}\right\}\\
{\vecstyle{Y}}&=&\mathrm{IFFT}\left\{\mathrm{FFT}\left\{w'_{1,i}\right\}^\ast\mathrm{FFT}\left\{w'_{2,i} \right\}\right\},
\end{eqnarray}
with the conjugate complex $\cdot ^\ast$ and 
where $\left\{w'_{1,i} \tilde x'_{1,i}\right\}$ and $\left\{w'_{1,i}\right\}$ are the zero-padded sets of weighted data values (after mean removal)
of the first data series and that of the weights respectively
and $\left\{w'_{2,i} \tilde x'_{2,i}\right\}$ and $\left\{w'_{2,i}\right\}$ those of the second data series and its appropriate weights.

This estimator has a similar systematic error as the variance estimator 
and the auto-covariance estimator above (see example in Fig.~\ref{fig:sim}c).
An appropriate estimation of the expectation of the cross-covariance function is
\begin{equation}
\rmE\{c_k\}=C_k+\varepsilon_k,
\label{eq:eckx}
\end{equation}
with the true cross-covariance function $C_k$ at lag time $\tau_k$ and the bias
\begin{eqnarray}
\varepsilon_k&=&\frac{\sum\limits_{i=0}^{N_1-1}\sum\limits_{j=0}^{N_2-1}w_{1,i} w_{2,j} C_{j-i}}{\left(\sum\limits_{i=0}^{N_1-1} w_{1,i}\right)\left(\sum\limits_{i=0}^{N_2-1} w_{2,i}\right)}
-\frac{\sum\limits_{i=\max(0,-k)}^{\min(N_1,N_2-k)-1}\sum\limits_{j=0}^{N_2-1}w_{1,i} w_{2,i+k} w_{2,j} C_{j-i}}
{\left(\sum\limits_{i=\max(0,-k)}^{\min(N_1,N_2-k)-1}w_{1,i} w_{2,i+k}\right)\left(\sum\limits_{i=0}^{N_2-1} w_{2,i}\right)}\nonumber\\
&&-\frac{\sum\limits_{i=\max(0,-k)}^{\min(N_1,N_2-k)-1}\sum\limits_{j=0}^{N_1-1}w_{1,i} w_{2,i+k} w_{1,j} C_{i+k-j}}
{\left(\sum\limits_{i=\max(0,-k)}^{\min(N_1,N_2-k)-1}w_{1,i} w_{2,i+k}\right)\left(\sum\limits_{i=0}^{N_1-1} w_{1,i}\right)},
\label{eq:eekx}
\end{eqnarray}
which is constant for uncorrelated data and only if the weights are identical for the two data sets, otherwise it varies with $k$.
The matrix ${\matstyle{A}}$, mapping a hypothetical covariance function ${\vecstyle{C}}$ onto the estimated one ${\vecstyle{c}}$ via
\begin{equation}
\rmE\{{\vecstyle{c}}\}={\matstyle{A}} {\vecstyle{C}},
\end{equation}
can be used to predict the estimated covariance at time lag $\tau_k$ as
\begin{equation}
\rmE\{c_k\}=\sum_{j=K_1}^{K_2}a_{kj} C_j
\end{equation}
with the elements $a_{kj}$ of the matrix ${\matstyle{A}}$.
The range $K_1\ldots K_2$ of covariances considered should include the full range of occurring correlations, such that 
all true covariance outside this interval can be neglected.

The elements of this matrix are\footnote{If all $w_i$ are constant, then the elements of this matrix become
\begin{eqnarray*}
a_{kj}&=&\delta_{k-j}-\frac{\min(N_1,N_2-j,N_2-k)-\max(0,-j,-k)}{N_2\left[\min(N_1,N_2-k)-\max(0,-k)\right]}\\
&&-\frac{\min\left[N_1,N_2-j,\max(0,N_1+k-j)\right]-\max\left[0,-j,\min(N_1,k-j)\right]}{N_1\left[\min(N_1,N_2-k)-\max(0,-k)\right]}\\
&&+\frac{\min(N_1,N_2-j)-\max(0,-j)}{N_1 N_2}\quad -N_1<j,k<N_2.
\end{eqnarray*}}
\begin{eqnarray}
a_{kj}&=&\delta_{k-j}
-\frac{\sum\limits_{i=\max(0,-j,-k)}^{\min(N_1,N_2-j,N_2-k)-1} w_{1,i} w_{2,i+j} w_{2,i+k}}{\left(\sum\limits_{i=\max(0,-k)}^{\min(N_1,N_2-k)-1} w_{1,i} w_{2,i+k}\right) \left(\sum\limits_{i=0}^{N_2-1} w_{2,i}\right)}\nonumber\\
&&\hspace{-10mm}-\frac{\sum\limits_{i=\max(0,-j,k-j)}^{\min(N_1,N_2-j,N_1+k-j)-1} w_{1,i} w_{2,i+j} w_{1,i+j-k}}{\left(\sum\limits_{i=\max(0,-k)}^{\min(N_1,N_2-k)-1} w_{1,i} w_{2,i+k}\right) \left(\sum\limits_{i=0}^{N_1-1} w_{1,i}\right)}
+\frac{\sum\limits_{i=\max(0,-j)}^{\min(N_1,N_2-j)-1} w_{1,i} w_{2,i+j}}{\left(\sum\limits_{i=0}^{N_1-1} w_{1,i}\right)\left(\sum\limits_{i=0}^{N_2-1} w_{2,i}\right)}\nonumber\\
\end{eqnarray}
again with
\begin{equation}
\delta_i=\left\{\begin{array}{ll}1&\mbox{for $i=0$}\\0&\mbox{otherwise}\end{array}\right.
\end{equation}
or
\begin{equation}
a_{kj}=\delta_{k-j}+\frac{Y_j}{\left(\sum\limits_{i=0}^{N_1-1} w_{1,i}\right)\left(\sum\limits_{i=0}^{N_2-1} w_{2,i}\right)}
-\frac{G_{kj}}{Y_k \left(\sum\limits_{i=0}^{N_2-1} w_{2,i}\right)}
-\frac{H_{kj}}{Y_k \left(\sum\limits_{i=0}^{N_1-1} w_{1,i}\right)}
\end{equation}
with
\begin{eqnarray}
{\vecstyle{G}}_k&=&\mathrm{IFFT}\left\{\mathrm{FFT}\left\{w'_{1,i}w'_{2,i+k} \right\}^\ast \mathrm{FFT}\left\{w'_{2,i} \right\}   \right\}\\
{\vecstyle{H}}_k&=&\mathrm{IFFT}\left\{\mathrm{FFT}\left\{w'_{1,i} \right\}^\ast \mathrm{FFT}\left\{w'_{2,i}w'_{1,i-k} \right\}    \right\},
\end{eqnarray}
involving again the (fast) discrete Fourier transform (FFT) and its inverse (IFFT).

The inverse of the matrix ${\matstyle{A}}^{-1}$ applied to the estimate ${\vecstyle{c}}$ yields an 
improved, bias-free estimate $\hat {\vecstyle{c}}$ of the cross-covariance
\begin{equation}
\hat {\vecstyle{c}}={\matstyle{A}}^{-1} {\vecstyle{c}}.
\end{equation}

For given $N_1$ samples $x_{1,i}$ and $N_2$ samples $x_{2,i}$, the covariance function after zero padding 
has $N_1+N_2-1$ non-zero values $c_k$ in the range $-(N_1-1)\ldots N_2-1$. 
Unfortunately, the appropriate matrix ${\matstyle{A}}$ then has some linear dependent equations and a direct 
inverse cannot be calculated. The inverse can be calculated only, if the covariance function is limited to 
the range $K_1\ldots K_2$ with $-(N_1-1)<K_1\le K_2<N_2-1$. The improved covariance estimate then is bias free, as long as 
the true covariance of the original signal is zero outside the reduced interval of lag times $\tau_{K_1}\ldots \tau_{K_2}$. 
This coincides with the requirement that the interval of investigated lag times is larger than the longest correlation lasts 
and the observation interval of the signal is at least a little longer than the largest lag time investigated.

\section{Numerical simulation}

\begin{figure}
\centerline{\includegraphics{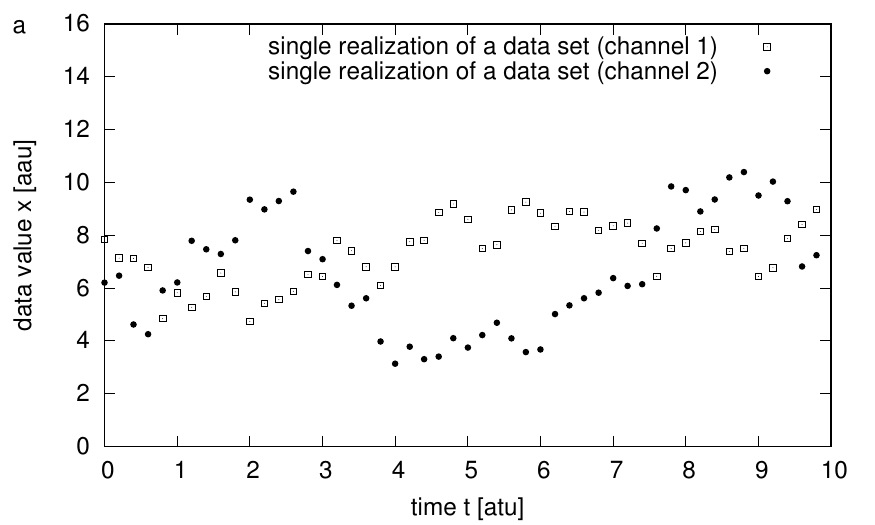}}\par
\centerline{\includegraphics{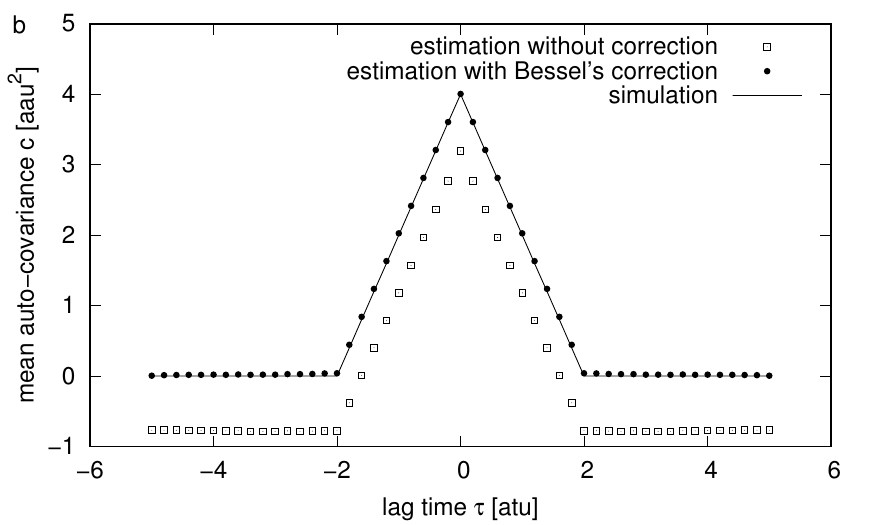}}\par
\centerline{\includegraphics{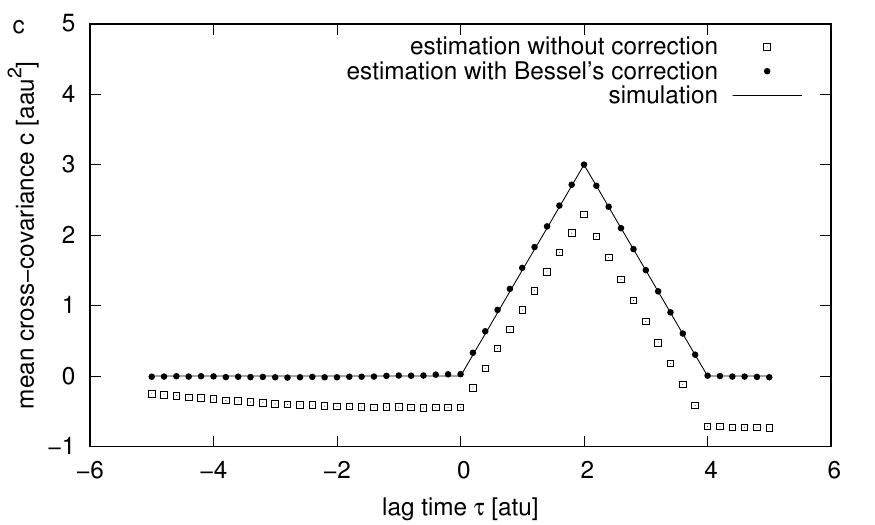}}\par
\caption{\label{fig:sim}a) Single realization of the data set from simulation.
b) Estimate of the auto-covariance function (empirical mean from 10\,000 realizations) 
without and with Bessel's correction for auto-covariance in comparison to the expected auto-covariance function according 
to the simulation process
c) Estimate of the cross-covariance function (empirical mean from 10\,000 realizations) 
without and with Bessel's correction for cross-covariance in comparison to the expected cross-covariance function according 
to the simulation process
($\unit{\sf atu}$ - arbitrary time unit, $\unit{\sf aau}$ - arbitrary amplitude unit)}
\end{figure}

To demonstrate the effect of Bessel's correction two linear random processes (moving average of order 10, all coefficients 0.1) 
with $\Delta t=\unit[0.2]{\sf atu}$ ($\unit{\sf atu}$ - arbitrary time unit) have been simulated,
each with a normal distribution with a variance of $\unit[4]{\sf aau^2}$ ($\unit{\sf aau}$ - arbitrary amplitude unit) 
and a mean of $\unit[8]{\sf aau}$. The two series 
have been coupled, yielding a cross-covariance of $\unit[3]{\sf aau^2}$ and one series has been time shifted to obtain a delay 
of $\unit[2]{\sf atu}$ between the two time series, which finally are limited to $N_1=N_2=50$ samples each. 
The weights have been random values from a uniform distribution between zero and one.
To obtain the empirical mean of the auto-covariance and the cross-covariance estimation, 10\,000 individual realizations 
(Fig.~\ref{fig:sim}a) have been simulated and analyzed (calculation of the mean values, mean removal and estimation of the 
auto-covariance function of one of the data sets and the cross-covariance function between the two data sets 
with $K_1=-25$ and $K_2=24$). Fig.~\ref{fig:sim}b and c compare the empirical mean of the auto-covariance estimate
and the cross-covariance estimates respectively without and with the proposed correction. Without the correction, 
the bias is obvious, all covariance values are underestimated here, while additionally 
a drift can be observed in the cross-covariance case, which in other cases may also lead to an over-estimation
at certain lag times. 
The introduced correction efficiently removes the bias and yields bias-free estimates of the auto-covariance function and
the cross-covariance function.

\section{Conclusion}

The removal of the estimated mean values from sampled, time-limited data sets
causes a bias in the estimates of the auto-covariance and the cross-covariance functions.
Based on the true covariance function, a prediction of the bias has been derived for 
such data sets with correlated samples including individual weighting of the samples. From the linear 
equations of the bias prediction an inverse matrix has been derived, which can be applied to the 
initial estimates of the covariance function to obtain an improved, bias-free estimate of the respective functions.
The corrected estimates then can be used to correct also the variance estimate 
in the case of correlated data. Numerical simulations have shown the improvements 
in estimating the covariance functions by the introduced procedures.

The findings well agree with the derivations of \citet{vogelsang_yang_16}, especially 
the linear dependencies of the respective system of equations and the feasibility 
of the inversion of an appropriate sub-matrix. The findings have been extended 
by the implementation of weighted averages in the estimation procedures, the 
investigation of the cross-covariance between different data sets,
the implementation of the fast Fourier transform to accelerate the calculations
and the bias-free estimation of the variance under the condition of correlated data samples. 

\section*{Acknowledgement}

The author gratefully acknowledges the fruitful discussion with Annette Witt.

\appendix

\section{Derivation of Eq.\ (\ref{eq:esq}) and (\ref{eq:varicorrect})}

From Eq.\ (\ref{eq:sq}) follows
\begin{eqnarray}
s^2&=&\frac{\sum\limits_{i=0}^{N-1} w_i \tilde x_i^2}{\sum\limits_{i=0}^{N-1} w_i}
=\frac{\sum\limits_{i=0}^{N-1} w_i \left(x_i-\bar x\right)^2}{\sum\limits_{i=0}^{N-1} w_i}\\
&=&\frac{\sum\limits_{i=0}^{N-1} w_i x_i^2}{\sum\limits_{i=0}^{N-1} w_i}
-2\frac{\sum\limits_{i=0}^{N-1} w_i x_i\bar x}{\sum\limits_{i=0}^{N-1} w_i}
+\frac{\sum\limits_{i=0}^{N-1} w_i \bar x^2}{\sum\limits_{i=0}^{N-1} w_i}\\
&=&\frac{\sum\limits_{i=0}^{N-1} w_i x_i^2}{\sum\limits_{i=0}^{N-1} w_i}
-2\frac{\sum\limits_{i=0}^{N-1} w_i x_i\left(\frac{\sum\limits_{j=0}^{N-1} w_j x_j}{\sum\limits_{j=0}^{N-1} w_j}\right)}{\sum\limits_{i=0}^{N-1} w_i}
+\frac{\sum\limits_{i=0}^{N-1} w_i \left(\frac{\sum\limits_{j=0}^{N-1} w_j x_j}{\sum\limits_{j=0}^{N-1} w_j}\right)^2}{\sum\limits_{i=0}^{N-1} w_i}\\
&=&\frac{\sum\limits_{i=0}^{N-1} w_i x_i^2}{\sum\limits_{i=0}^{N-1} w_i}
-2\frac{\sum\limits_{i=0}^{N-1}\sum\limits_{j=0}^{N-1} w_i w_j x_i x_j}{\left(\sum\limits_{i=0}^{N-1} w_i\right)^2}
+\left(\frac{\sum\limits_{j=0}^{N-1} w_j x_j}{\sum\limits_{j=0}^{N-1} w_j}\right)^2\\
&=&\frac{\sum\limits_{i=0}^{N-1} w_i x_i^2}{\sum\limits_{i=0}^{N-1} w_i}
-2\frac{\sum\limits_{i=0}^{N-1}\sum\limits_{j=0}^{N-1} w_i w_j x_i x_j}{\left(\sum\limits_{i=0}^{N-1} w_i\right)^2}
+\frac{\sum\limits_{i=0}^{N-1}\sum\limits_{j=0}^{N-1} w_i w_j x_i x_j}{\left(\sum\limits_{i=0}^{N-1} w_i\right)^2}\\
&=&\frac{\sum\limits_{i=0}^{N-1} w_i x_i^2}{\sum\limits_{i=0}^{N-1} w_i}
-\frac{\sum\limits_{i=0}^{N-1}\sum\limits_{j=0}^{N-1} w_i w_j x_i x_j}{\left(\sum\limits_{i=0}^{N-1} w_i\right)^2}\\
&=&\frac{\sum\limits_{i=0}^{N-1}\sum\limits_{j=0}^{N-1} w_i w_j \left( x_i^2 - x_i x_j\right)}{\left(\sum\limits_{i=0}^{N-1} w_i\right)^2}.
\end{eqnarray}
The expectation of $s^2$ then is
\begin{eqnarray}
\rmE\{s^2\}&=&\frac{\sum\limits_{i=0}^{N-1}\sum\limits_{j=0}^{N-1} w_i w_j \left[ \left( \sigma_x^2 +\mu^2\right)- \left( C_{j-i}+\mu^2\right)\right]}{\left(\sum\limits_{i=0}^{N-1} w_i\right)^2}\\
&=&\frac{\sum\limits_{i=0}^{N-1}\sum\limits_{j=0}^{N-1} w_i w_j \sigma_x^2}{\left(\sum\limits_{i=0}^{N-1} w_i\right)^2}-\frac{\sum\limits_{i=0}^{N-1}\sum\limits_{j=0}^{N-1} w_i w_j C_{j-i}}{\left(\sum\limits_{i=0}^{N-1} w_i\right)^2}\\
&=&\sigma_x^2-\sigma_{\bar x}^2.
\end{eqnarray}

\section{Derivation of Eqs.\ (\ref{eq:eck}) and (\ref{eq:eek})}

From Eq.\ (\ref{eq:ck}) follows
\begin{eqnarray}
c_k&=&\frac{\sum\limits_{i=\max(0,-k)}^{\min(N,N-k)-1} w_i w_{i+k} \tilde x_i \tilde x_{i+k}}{\sum\limits_{i=\max(0,-k)}^{\min(N,N-k)-1} w_i w_{i+k}}\\
&=&\frac{\sum\limits_{i=\max(0,-k)}^{\min(N,N-k)-1} w_i w_{i+k} \left(x_i - \bar x\right) \left(x_{i+k} - \bar x\right)}{\sum\limits_{i=\max(0,-k)}^{\min(N,N-k)-1} w_i w_{i+k}}\\
&=&\frac{\sum\limits_{i=\max(0,-k)}^{\min(N,N-k)-1} w_i w_{i+k} x_i x_{i+k}}{\sum\limits_{i=\max(0,-k)}^{\min(N,N-k)-1} w_i w_{i+k}}
-\bar x\frac{\sum\limits_{i=\max(0,-k)}^{\min(N,N-k)-1} w_i w_{i+k} \left(x_i+x_{i+k}\right)}{\sum\limits_{i=\max(0,-k)}^{\min(N,N-k)-1} w_i w_{i+k}}
+\bar x^2\nonumber\\
\\
&=&\frac{\sum\limits_{i=\max(0,-k)}^{\min(N,N-k)-1} w_i w_{i+k} x_i x_{i+k}}{\sum\limits_{i=\max(0,-k)}^{\min(N,N-k)-1} w_i w_{i+k}}
\nonumber\\
&&-\left(\frac{\sum\limits_{j=0}^{N-1} w_j x_j}{\sum\limits_{j=0}^{N-1} w_j}\right)\frac{\sum\limits_{i=\max(0,-k)}^{\min(N,N-k)-1} w_i w_{i+k} \left(x_i+x_{i+k}\right)}{\sum\limits_{i=\max(0,-k)}^{\min(N,N-k)-1} w_i w_{i+k}}
+\left(\frac{\sum\limits_{j=0}^{N-1} w_j x_j}{\sum\limits_{j=0}^{N-1} w_j}\right)^2\\
&=&\frac{\sum\limits_{i=\max(0,-k)}^{\min(N,N-k)-1} w_i w_{i+k} x_i x_{i+k}}{\sum\limits_{i=\max(0,-k)}^{\min(N,N-k)-1} w_i w_{i+k}}
\nonumber\\
&&-\frac{\sum\limits_{i=\max(0,-k)}^{\min(N,N-k)-1} \sum\limits_{j=0}^{N-1} w_i w_{i+k} w_j \left(x_i+x_{i+k}\right)x_j}{\left(\sum\limits_{i=\max(0,-k)}^{\min(N,N-k)-1} w_i w_{i+k}\right)\left(\sum\limits_{j=0}^{N-1} w_j\right)}
+\frac{\sum\limits_{i=0}^{N-1} \sum\limits_{j=0}^{N-1} w_i w_j x_i x_j}{\left(\sum\limits_{j=0}^{N-1} w_j\right)^2}\\
&=&\frac{\sum\limits_{i=\max(0,-k)}^{\min(N,N-k)-1} w_i w_{i+k} x_i x_{i+k}}{\sum\limits_{i=\max(0,-k)}^{\min(N,N-k)-1} w_i w_{i+k}}+\frac{\sum\limits_{i=0}^{N-1} \sum\limits_{j=0}^{N-1} w_i w_j x_i x_j}{\left(\sum\limits_{i=0}^{N-1} w_i\right)^2}
\nonumber\\
&&-\frac{\sum\limits_{i=\max(0,-k)}^{\min(N,N-k)-1} \sum\limits_{j=0}^{N-1} w_i w_{i+k} w_j \left(x_i+x_{i+k}\right)x_j}{\left(\sum\limits_{i=\max(0,-k)}^{\min(N,N-k)-1} w_i w_{i+k}\right)\left(\sum\limits_{i=0}^{N-1} w_i\right)}.
\end{eqnarray}
The expectation of $c_k$ then is
\begin{eqnarray}
\rmE\{c_k\}&=&\frac{\sum\limits_{i=\max(0,-k)}^{\min(N,N-k)-1} w_i w_{i+k} \left(C_k+\mu^2\right)}{\sum\limits_{i=\max(0,-k)}^{\min(N,N-k)-1} w_i w_{i+k}}
+\frac{\sum\limits_{i=0}^{N-1} \sum\limits_{j=0}^{N-1} w_i w_j \left(C_{j-i}+\mu^2\right)}{\left(\sum\limits_{i=0}^{N-1} w_i\right)^2}
\nonumber\\
&&-\frac{\sum\limits_{i=\max(0,-k)}^{\min(N,N-k)-1} \sum\limits_{j=0}^{N-1} w_i w_{i+k} w_j \left(C_{j-i}+C_{j-(i+k)}+2\mu^2\right)}{\left(\sum\limits_{i=\max(0,-k)}^{\min(N,N-k)-1} w_i w_{i+k}\right)\left(\sum\limits_{i=0}^{N-1} w_i\right)}\\
&=&C_k
+\frac{\sum\limits_{i=0}^{N-1} \sum\limits_{j=0}^{N-1} w_i w_j C_{j-i}}{\left(\sum\limits_{i=0}^{N-1} w_i\right)^2}
\nonumber\\
&&-\frac{\sum\limits_{i=\max(0,-k)}^{\min(N,N-k)-1} \sum\limits_{j=0}^{N-1} w_i w_{i+k} w_j \left(C_{j-i}+C_{i+k-j}\right)}{\left(\sum\limits_{i=\max(0,-k)}^{\min(N,N-k)-1} w_i w_{i+k}\right)\left(\sum\limits_{i=0}^{N-1} w_i\right)}\\
&=&C_k
+\sigma_{\bar x}^2
-\frac{\sum\limits_{i=\max(0,-k)}^{\min(N,N-k)-1} \sum\limits_{j=0}^{N-1} w_i w_{i+k} w_j \left(C_{j-i}+C_{i+k-j}\right)}{\left(\sum\limits_{i=\max(0,-k)}^{\min(N,N-k)-1} w_i w_{i+k}\right)\left(\sum\limits_{i=0}^{N-1} w_i\right)}.
\end{eqnarray}

\section{Derivation of Eqs.\ (\ref{eq:eckx}) and (\ref{eq:eekx})}

From Eq.\ (\ref{eq:ckx}) follows
\begin{eqnarray}
c_k&=&\frac{\sum\limits_{i=\max(0,-k)}^{\min(N_1,N_2-k)-1} w_{1,i} w_{2,i+k} \tilde x_{1,i} \tilde x_{2,i+k}}{\sum\limits_{i=\max(0,-k)}^{\min(N_1,N_2-k)-1} w_{1,i} w_{2,i+k}}\\
&=&\frac{\sum\limits_{i=\max(0,-k)}^{\min(N_1,N_2-k)-1} w_{1,i} w_{2,i+k} \left( x_{1,i} - \bar x_1 \right) \left( x_{2,i+k} - \bar x_2\right)}{\sum\limits_{i=\max(0,-k)}^{\min(N_1,N_2-k)-1} w_{1,i} w_{2,i+k}}\\
&=&\frac{\sum\limits_{i=\max(0,-k)}^{\min(N_1,N_2-k)-1} w_{1,i} w_{2,i+k} x_{1,i} x_{2,i+k}}{\sum\limits_{i=\max(0,-k)}^{\min(N_1,N_2-k)-1} w_{1,i} w_{2,i+k}}
-\bar x_2\frac{\sum\limits_{i=\max(0,-k)}^{\min(N_1,N_2-k)-1} w_{1,i} w_{2,i+k} x_{1,i}}{\sum\limits_{i=\max(0,-k)}^{\min(N_1,N_2-k)-1} w_{1,i} w_{2,i+k}}
\nonumber\\
&&-\bar x_1\frac{\sum\limits_{i=\max(0,-k)}^{\min(N_1,N_2-k)-1} w_{1,i} w_{2,i+k} x_{2,i+k}}{\sum\limits_{i=\max(0,-k)}^{\min(N_1,N_2-k)-1} w_{1,i} w_{2,i+k}}
+\bar x_1 \bar x_2\\
&=&\frac{\sum\limits_{i=\max(0,-k)}^{\min(N_1,N_2-k)-1} w_{1,i} w_{2,i+k} x_{1,i} x_{2,i+k}}{\sum\limits_{i=\max(0,-k)}^{\min(N_1,N_2-k)-1} w_{1,i} w_{2,i+k}}
\nonumber\\
&&-\left(\frac{\sum\limits_{j=0}^{N_2-1} w_{2,j} x_{2,j}}{\sum\limits_{j=0}^{N_2-1} w_{2,j}}\right)\frac{\sum\limits_{i=\max(0,-k)}^{\min(N_1,N_2-k)-1} w_{1,i} w_{2,i+k} x_{1,i}}{\sum\limits_{i=\max(0,-k)}^{\min(N_1,N_2-k)-1} w_{1,i} w_{2,i+k}}
\nonumber\\
&&-\left(\frac{\sum\limits_{j=0}^{N_1-1} w_{1,j} x_{1,j}}{\sum\limits_{j=0}^{N_1-1} w_{1,j}}\right)\frac{\sum\limits_{i=\max(0,-k)}^{\min(N_1,N_2-k)-1} w_{1,i} w_{2,i+k} x_{2,i+k}}{\sum\limits_{i=\max(0,-k)}^{\min(N_1,N_2-k)-1} w_{1,i} w_{2,i+k}}
\nonumber\\
&&+\left(\frac{\sum\limits_{j=0}^{N_1-1} w_{1,j} x_{1,j}}{\sum\limits_{j=0}^{N_1-1} w_{1,j}}\right)\left(\frac{\sum\limits_{j=0}^{N_2-1} w_{2,j} x_{2,j}}{\sum\limits_{j=0}^{N_2-1} w_{2,j}}\right)\\
&=&\frac{\sum\limits_{i=\max(0,-k)}^{\min(N_1,N_2-k)-1} w_{1,i} w_{2,i+k} x_{1,i} x_{2,i+k}}{\sum\limits_{i=\max(0,-k)}^{\min(N_1,N_2-k)-1} w_{1,i} w_{2,i+k}}
\nonumber\\
&&-\frac{\sum\limits_{i=\max(0,-k)}^{\min(N_1,N_2-k)-1} \sum\limits_{j=0}^{N_2-1} w_{1,i} w_{2,i+k} w_{2,j} x_{1,i} x_{2,j}}{\left(\sum\limits_{i=\max(0,-k)}^{\min(N_1,N_2-k)-1} w_{1,i} w_{2,i+k}\right)\left(\sum\limits_{j=0}^{N_2-1} w_{2,j}\right)}
\nonumber\\
&&-\frac{\sum\limits_{i=\max(0,-k)}^{\min(N_1,N_2-k)-1} \sum\limits_{j=0}^{N_1-1} w_{1,i} w_{2,i+k} w_{1,j} x_{2,i+k} x_{1,j}}{\left(\sum\limits_{i=\max(0,-k)}^{\min(N_1,N_2-k)-1} w_{1,i} w_{2,i+k}\right)\left(\sum\limits_{j=0}^{N_1-1} w_{1,j}\right)}
\nonumber\\
&&+\frac{\sum\limits_{i=0}^{N_1-1} \sum\limits_{j=0}^{N_2-1} w_{1,i} w_{2,j} x_{1,i} x_{2,j}}{\left(\sum\limits_{i=0}^{N_1-1} w_{1,i}\right)\left(\sum\limits_{j=0}^{N_2-1} w_{2,j}\right)}\\
&=&\frac{\sum\limits_{i=\max(0,-k)}^{\min(N_1,N_2-k)-1} w_{1,i} w_{2,i+k} x_{1,i} x_{2,i+k}}{\sum\limits_{i=\max(0,-k)}^{\min(N_1,N_2-k)-1} w_{1,i} w_{2,i+k}}
+\frac{\sum\limits_{i=0}^{N_1-1} \sum\limits_{j=0}^{N_2-1} w_{1,i} w_{2,j} x_{1,i} x_{2,j}}{\left(\sum\limits_{i=0}^{N_1-1} w_{1,i}\right)\left(\sum\limits_{i=0}^{N_2-1} w_{2,i}\right)}
\nonumber\\
&&-\frac{\sum\limits_{i=\max(0,-k)}^{\min(N_1,N_2-k)-1} \sum\limits_{j=0}^{N_2-1} w_{1,i} w_{2,i+k} w_{2,j} x_{1,i} x_{2,j}}{\left(\sum\limits_{i=\max(0,-k)}^{\min(N_1,N_2-k)-1} w_{1,i} w_{2,i+k}\right)\left(\sum\limits_{i=0}^{N_2-1} w_{2,i}\right)}
\nonumber\\
&&-\frac{\sum\limits_{i=\max(0,-k)}^{\min(N_1,N_2-k)-1} \sum\limits_{j=0}^{N_1-1} w_{1,i} w_{2,i+k} w_{1,j} x_{1,j} x_{2,i+k}}{\left(\sum\limits_{i=\max(0,-k)}^{\min(N_1,N_2-k)-1} w_{1,i} w_{2,i+k}\right)\left(\sum\limits_{i=0}^{N_1-1} w_{1,i}\right)}.
\end{eqnarray}
The expectation of $c_k$ then is
\begin{eqnarray}
\rmE\{c_k\}&=&\frac{\sum\limits_{i=\max(0,-k)}^{\min(N_1,N_2-k)-1} w_{1,i} w_{2,i+k} \left(C_k+\mu_1 \mu_2\right)}{\sum\limits_{i=\max(0,-k)}^{\min(N_1,N_2-k)-1} w_{1,i} w_{2,i+k}}
\nonumber\\
&&+\frac{\sum\limits_{i=0}^{N_1-1} \sum\limits_{j=0}^{N_2-1} w_{1,i} w_{2,j} \left(C_{j-i}+\mu_1 \mu_2\right)}{\left(\sum\limits_{i=0}^{N_1-1} w_{1,i}\right)\left(\sum\limits_{i=0}^{N_2-1} w_{2,i}\right)}
\nonumber\\
&&-\frac{\sum\limits_{i=\max(0,-k)}^{\min(N_1,N_2-k)-1} \sum\limits_{j=0}^{N_2-1} w_{1,i} w_{2,i+k} w_{2,j} \left(C_{j-i}+\mu_1 \mu_2\right)}{\left(\sum\limits_{i=\max(0,-k)}^{\min(N_1,N_2-k)-1} w_{1,i} w_{2,i+k}\right)\left(\sum\limits_{i=0}^{N_2-1} w_{2,i}\right)}
\nonumber\\
&&-\frac{\sum\limits_{i=\max(0,-k)}^{\min(N_1,N_2-k)-1} \sum\limits_{j=0}^{N_1-1} w_{1,i} w_{2,i+k} w_{1,j} \left(C_{i+k-j}+\mu_1 \mu_2\right)}{\left(\sum\limits_{i=\max(0,-k)}^{\min(N_1,N_2-k)-1} w_{1,i} w_{2,i+k}\right)\left(\sum\limits_{i=0}^{N_1-1} w_{1,i}\right)}\\
&=&C_k
+\frac{\sum\limits_{i=0}^{N_1-1} \sum\limits_{j=0}^{N_2-1} w_{1,i} w_{2,j} C_{j-i}}{\left(\sum\limits_{i=0}^{N_1-1} w_{1,i}\right)\left(\sum\limits_{i=0}^{N_2-1} w_{2,i}\right)}
\nonumber\\
&&-\frac{\sum\limits_{i=\max(0,-k)}^{\min(N_1,N_2-k)-1} \sum\limits_{j=0}^{N_2-1} w_{1,i} w_{2,i+k} w_{2,j} C_{j-i}}{\left(\sum\limits_{i=\max(0,-k)}^{\min(N_1,N_2-k)-1} w_{1,i} w_{2,i+k}\right)\left(\sum\limits_{i=0}^{N_2-1} w_{2,i}\right)}
\nonumber\\
&&-\frac{\sum\limits_{i=\max(0,-k)}^{\min(N_1,N_2-k)-1} \sum\limits_{j=0}^{N_1-1} w_{1,i} w_{2,i+k} w_{1,j} C_{i+k-j}}{\left(\sum\limits_{i=\max(0,-k)}^{\min(N_1,N_2-k)-1} w_{1,i} w_{2,i+k}\right)\left(\sum\limits_{i=0}^{N_1-1} w_{1,i}\right)}.
\end{eqnarray}

\bibliography{BC4corr}

\end{document}